\documentclass{aa}  

\usepackage{graphicx}
\usepackage{txfonts}
\usepackage{lipsum}
\usepackage{subcaption}
\usepackage{lscape}
\usepackage{placeins}
\usepackage{ulem}
\usepackage{amsmath}
\usepackage{booktabs}
\usepackage[dvipsnames]{xcolor}
\usepackage[colorlinks=true, linkcolor=blue, citecolor=blue]{hyperref}

\newcommand{\etaI}{\ensuremath{\eta^\mathrm{I}}}
\newcommand{\etaPI}{\ensuremath{\eta^\mathrm{PI}}}
\newcommand{\zetadisk}{\ensuremath{\zeta_{\mathrm{disk}}}}
\newcommand{\zetaeff}{\ensuremath{\zeta_{\mathrm{eff}}}}

\begin{document}

   \title{The colors of protoplanetary disks}
   \subtitle{Impact of nonspherical dust particles on the polarized and total intensity colors}

   \author{M. Roumesy\inst{1}
          \and
          J. Ma\inst{1}
          \and
          R. Tazaki\inst{2}
          \and
          F. Ménard\inst{1}
          \and
          G. Duchêne\inst{1, 3}
          }
   \institute{Univ. Grenoble Alpes, CNRS, IPAG, 38000 Grenoble, France
   \and Department of Earth Science and Astronomy, The University of Tokyo, Tokyo 153-8902, Japan
   \and Astronomy Department, University of California Berkeley, Berkeley CA 94720-3411, USA}
   \date{Received --; accepted --}

\abstract
{Dust properties in protoplanetary disks remain poorly understood despite their central role in planetary formation. A detailed observational characterization of their properties is crucial to improving our understanding of grain growth. Recent studies, based on analyses of scattered-light images and scattering phase functions, provide promising insights into dust characterization. However, the phase function alone is insufficient to constrain dust properties efficiently. In particular, degeneracies between dust size, composition, and morphology remain a major obstacle to linking observations to realistic dust models.}
{We investigate the behavior of the disk color as a function of dust size, shape, and composition. The goal is to combine this with phase function measurements to most effectively constrain the dust properties.}
{We used the 3D radiative transfer code {\tt{MCFOST}} to generate synthetic scattered-light images of protoplanetary disks from the visible to the near-infrared, in both total and polarized intensity. We considered compact spherical grains using Mie theory. We used the optical properties computed with the T-matrix method or the discrete dipole approximation for irregular particles and for compact and fractal aggregates from the {\tt{AggScatVIR}} database. We measured the disk colors directly from the synthetic images.}
{We obtain five main results.
(i) For aggregates larger than the wavelength, the monomer size plays a crucial role in determining the colors, whereas the aggregate size has a negligible effect, especially for low fractal dimensions.
(ii) The polarized color is more sensitive to the dust properties than the total intensity color, likely because it combines the wavelength dependences of both the scattered intensity and the degree of polarization.
(iii) The effective medium theory is a reliable approximation only for relatively compact aggregates composed of small monomers and becomes inaccurate for highly porous aggregates or those composed of larger monomers.
(iv) The effective albedo is useful for predicting disk colors, but radiative transfer calculations are still required to account for the multiple scattering effect and quantitatively interpret disk colors.
(v) For HD~100546, the red optical to near-infrared colors exclude compact grains and organic-rich compositions but are compatible with fractal or compact aggregates of amorphous carbon composition containing large monomers (200--400~nm).}
{Our study of color based on synthetic data and its application to the example of the disk around HD~100546 shows that combining color measurements, or a color measurement with a polarized scattering phase function, offers a promising approach to breaking the degeneracy between dust models. Extending this approach to larger samples of disks with multiwavelength observations should provide better constraints on the properties and evolution of dust and motivate the use of color in future studies of protoplanetary disks.}

\keywords{polarization, scattering, protoplanetary disks}
\maketitle
\nolinenumbers

\section{Introduction} \label{Sect:Introduction}
\begin{figure*}[ht]
    \centering
    \includegraphics[width=0.99\linewidth]{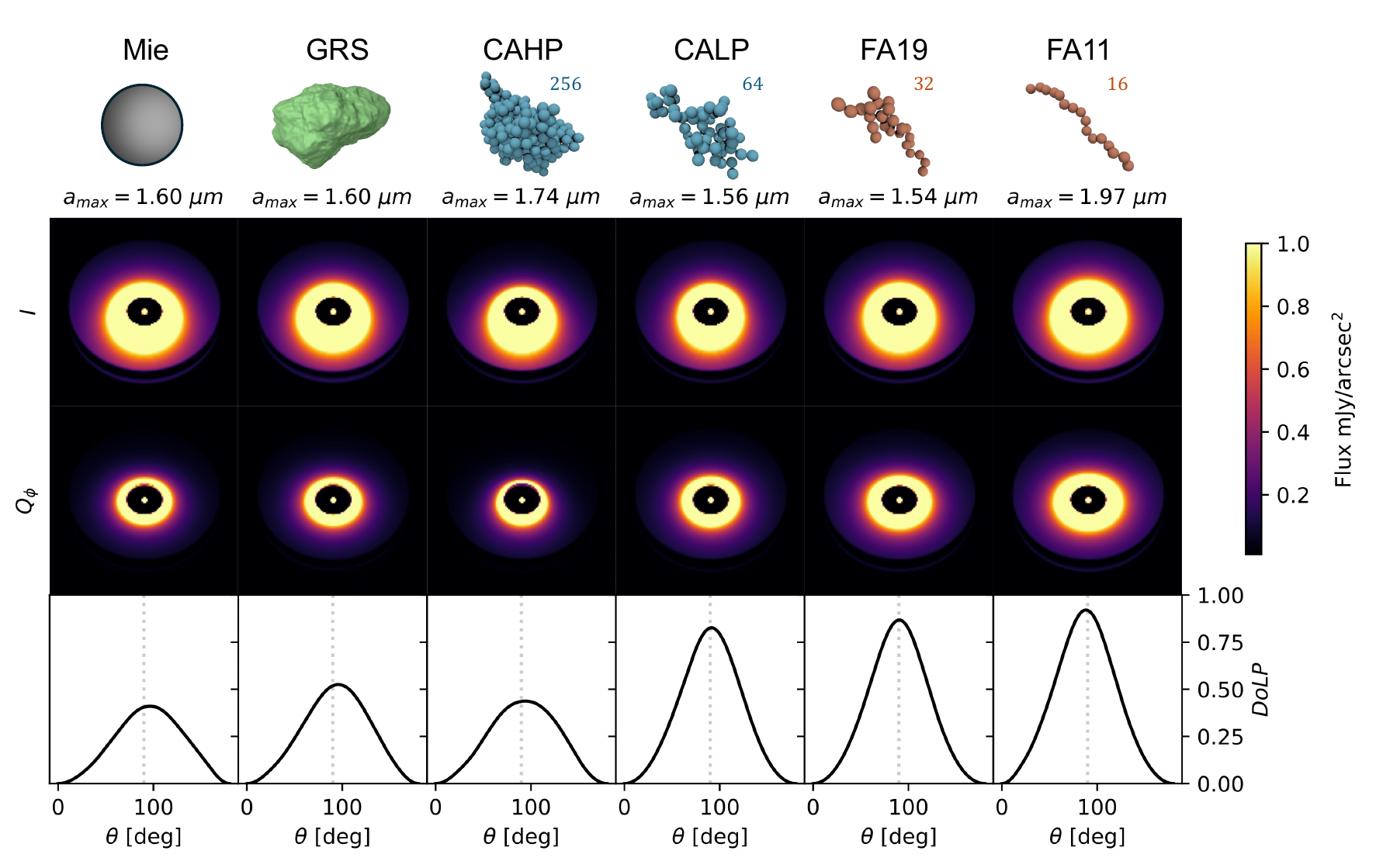}
    \caption{Synthetic images for the standard disk model, inclined at $30^\circ$, observed for total (\textit{top}) and polarized (\textit{bottom}) intensity at $1.63~\mu$m. Disks are composed of different dust models, as noted for each column. For the aggregates, we fix the monomer size to $200$ nm and vary the number of monomers to recover a maximum characteristic radius ($a_\mathrm{c, max}$) that is comparable to the observational wavelength. We also provide the number of monomers next to the aggregate images.}
    \label{fig:standarddisk}
\end{figure*}
The dust particles in protoplanetary disks are key actors of the planet formation process, earning these disks the title of ``planetary birthplaces'' \citep{keppler_discovery_2018, pinte_kinematic_2018, teague_kinematical_2018}. Dust grains evolve under the influence of complex coagulation and fragmentation mechanisms \citep{Birnstiel2024}, which affect their size, internal structure, and chemical composition. These processes, driven by hydrodynamic interactions such as vertical settling and radial drift \citep{villenave_turbulence_2025, villenave_introduction_2025}, modify the distribution of dust within the disk. The early stages of this evolution remain poorly understood, however, particularly how millimeter-sized grains grow into planetesimals within a very short timescale \citep{drazkowska_planet_2022}.

With major progress in high-resolution imaging in infrared scattered light, recent observations of protoplanetary disks \citep{benisty_optical_2022, valegard_disk_2024, ginski_sphere_2024, garufi_sphere_2024} have paved the way for an in-depth study of the geometry of these systems \citep{min_effects_2012, ma_model_2022}. Above all, light scattering is extremely sensitive to the physical properties of dust grains, in particular their size, composition, and shape \citep{kimura_light_2006, shen_modeling_2008, shen_modeling_2009, tazaki_light_2016, Min2016, tazaki_light_2018, halder_dependence_2018, ysard_optical_2018}, providing insights into the nature of the dust particles evolving in protoplanetary disks.

Recent scattered-light observations of protoplanetary disks have highlighted both the usefulness and the limitations of using scattering phase functions to constrain dust properties. The scattering phase function describes the angular dependence of scattered light and is sensitive to the size, composition, and morphology of dust particles \citep[see the review by][]{Potapov2025}. High-quality scattered-light images of protoplanetary disks now allow us to extract their scattering phase functions \citep{Pinte2007, stolker_scattered_2016, ginski_direct_2016, ginski_observed_2023, tazaki_fractal_2023, chen_multiband_2024, roumesy_dragys_2025, roumesy2026}. However, \citet{roumesy2026} show that the phase function alone often leads to strong degeneracies among dust parameters.
Complementary constraints are therefore required, and multiwavelength scattered-light observations can provide additional information on the dust-scattering properties.
Several studies in recent years have measured disk colors in terms of both total and polarized intensity \citep[e.g.,][]{fukagawa_subaru_2010, avenhaus_disks_2018, ma_color_2024}. In the near-infrared, most disks exhibit approximately gray scattered light in total intensity \citep[e.g.,][]{fukagawa_subaru_2010}, although some disks show red colors \citep[e.g.,][]{krist_hubble_2005, mulders_why_2013}. Near-infrared polarized-intensity colors are also often approximately gray, while red colors have been reported for some disks, such as IM Lup \citep{avenhaus_disks_2018}. In contrast, more recent optical imaging reveals a predominance of reddish colors in polarized intensity \citep{hunziker_hd_2021, tschudi_quantitative_2021, ma_quantitative_2023, stolker_shadows_2016, stolker_variable_2017, monnier_multiple_2019, ma_color_2024}.

The disk color in total and/or polarized intensity is governed by the dust properties and is therefore sensitive to the grain size distribution, as well as to grain composition and structure \citep[e.g.,][]{Whitney1995, draine_scattering_2003, Golimowski2006, Ardila2007, Debes2008, Min2016, tazaki_effect_2019}. As an example, small grains tend to produce blue scattered light due to Rayleigh scattering, whereas larger grains tend to produce gray or red scattered light. In the specific case of HD~100546, \citet{mulders_why_2013} showed, using the distribution of hollow spheres (DHS) method, that particles larger than several microns can reproduce its remarkably red color. The effect of a more complex particle morphology on disk colors has been explored in several theoretical studies, including the work of \cite{tazaki_effect_2019}, who used the T-matrix light-scattering method to model scattering by aggregates. They found that highly porous fractal aggregates do not produce red disk colors, in contrast to compact grains, even if their sizes are much larger than the wavelength. This result indicates that disk color depends not only on grain size but also on the structure and porosity of dust particles. However, \cite{tazaki_effect_2019} explored only a limited parameter space in terms of aggregate size, composition, and structure. A systematic exploration of these parameters and their effects on disk colors has yet to be performed.

In this study, we investigated the optical and near-infrared colors of protoplanetary disks in total and polarized intensity for a broad range of dust models, from compact spherical grains to irregular particles and compact or fractal aggregates. In addition to the commonly used Mie and DHS approximations, the calculations rely on the optical properties of nonspherical particles available in the {\tt{AggScatVIR}} database, which were computed using T-matrix and discrete-dipole approximation light-scattering computations \citep{tazaki_size_2022, tazaki_fractal_2023}. We incorporated the corresponding wavelength-dependent Mueller matrices into the 3D radiative transfer tool {\tt{MCFOST}} to produce synthetic disk images. Section \ref{Sect:Models} describes the fiducial disk model and the complex dust models investigated, as well as the method for incorporating them into {\tt{MCFOST}}. We show the disk color results in Sect. \ref{Sect:Results}, where we also discuss the impact of dust properties, and especially the role of monomer size. In Sect. \ref{Sect:Discussion} we discuss the results and the use of effective albedo as a color proxy, and we apply our framework to the specific case of HD~100546. Finally, a general conclusion on the color of the disks is presented in Sect. \ref{Sect:Conclusion}.

\section{Models} \label{Sect:Models}
\subsection{Disk model} \label{Sect:Diskmodel}
We performed Monte Carlo radiative transfer computations with {\tt{MCFOST}} \citep{pinte_monte_2006, pinte_benchmark_2009} to construct synthetic images of protoplanetary disks in total and polarized intensity. The fiducial disk model comprises a typical young solar-mass pre-main-sequence star surrounded by an inner disk from $0.1$ to $5$ au with a dust mass of $10^{-9}M_\odot$. The main disk extends from $50$ to $200$ au, aligned with the inner disk, with a total dust mass of $10^{-5}M_\odot$. The gas pressure scale height follows the power law $h(r) = h_0 (r/r_0)^{\beta}$ with $h_0 = 5$au at $r_0=100$au and a flaring of $\beta=1.125$. The disk lies at $100$~pc, and we fixed the disk inclination to $30$\degr. The resulting images at different wavelengths from visible to near-infrared are composed of $201\times201$ pixels, providing a field of view of $\sim5^"\times5^"$. We simulated synthetic disk images for each of these dust models (see Sect. \ref{Sect:Dustmodel}) observed at $0.554\mu$m, $0.735\mu$m, $1.25\mu$m, $1.63\mu$m, and $2.18\mu$m.

\subsection{Dust models} \label{Sect:Dustmodel}
To investigate the influence of dust morphology on the colors of protoplanetary disks, we considered several dust models in our {\tt{MCFOST}} simulations. Our reference model consists of compact spherical grains, whose optical properties are calculated using Mie theory \citep{Mie1908, bohren_absorption_1983}. We also considered the DHS model \citep{min_shape_2003, min_modeling_2005, min_optical_2009}, designed to approximate the scattering properties of complex particles as aggregates. We adopted the commonly used maximum void-volume fraction of $V_\mathrm{max}=0.8$.

For more complex dust grains, we used the {\tt{AggScatVIR}} (Aggregate Scattering for Visible and Infrared wavelengths) database constructed by \cite{tazaki_size_2022} and \cite{tazaki_fractal_2023}, which provides the optical properties of nonspherical particles. It includes three classes of particles. 
First, irregular grains are represented by Gaussian random spheres (GRS), with maximum grain sizes between $0.2$ and $1.6~\mu\mathrm{m}$. 
Second, the database contains compact aggregates composed of monomers with radii of $100$, $200$, or $400~$nm, which reach characteristic radii of up to $3~\mu$m. These particles are divided into the CAHP, CAMP, and CALP families, corresponding to high-, medium-, and low-porosity aggregates, respectively. Their porosity is defined by the volume fraction of void within the total aggregate volume.
Finally, fractal aggregates are composed of monomers with radii of  $100$, $150$, $200$, $300$ and $400~\mathrm{nm}$, and the characteristic radii can reach $\sim12~\mu$m. These particles are characterized by their fractal dimension $D_f$, which quantifies how the dust mass (i.e., the number of monomers) scales with the aggregate size, $N \propto (a_g/a_\mathrm{mon})^{D_f}$, where $N$ is the number of monomers, $a_\mathrm{mon}$ is the monomer radius and $a_g$ is the radius of gyration. A lower $D_f$ corresponds to a more open and porous structure, whereas a larger $D_f$ indicates a more compact aggregate. The {\tt{AggScatVIR}} database provides four models: FA11, FA13, FA15, and FA19, which correspond to $D_f = 1.1, 1.3, 1.5,$ and $1.9,$ respectively. 

A summary of all dust models and their available size ranges is given in Appendix~\ref{APP:DustModels} and Table~\ref{tab:dustmodels}. For every {\tt{AggScatVIR}} model, as well as Mie and DHS models, we used two different compositions made of a mixture of water ice (20\% in mass), silicate ($32.9\%$), and troilite ($7.4\%$), adding $39.7\%$ of organic matter (hereafter ``org''; \citealt{henning_dust_1996}) or amorphous carbon (hereafter ``amc''; \citealt{zubko_optical_1996}). 

In the following, we present the results obtained for a representative subset of dust models, selected to clarify the results on the disk color (from least to most porous: Mie, DHS, GRS, CALP, CAHP, FA19, and FA11) considering a grain size distribution and amorphous carbon (amc) composition. This selection highlights the main effects of grain shape and monomer size while keeping the comparison tractable. Figure~\ref{fig:standarddisk} provides an overview of synthetic scattered-light images for the selected subset of dust models, including total intensity $I$, polarized intensity $Q_\phi$ and degree of linear polarization (DoLP). It shows that the aggregate models generally yield higher DoLP than the compact-particle models, illustrating the sensitivity of polarized scattered light to the internal structure of the dust particles.

We describe the size distribution of all dust models using the volume-equivalent radius $a_\mathrm{V}$. This quantity is defined as the radius of a compact sphere containing the same material volume as the considered particle. We adopted a power-law size distribution following the standard MRN prescription \citep{Mathis1977},
\begin{equation}
    n(a_\mathrm{V})\mathrm{d}a_\mathrm{V} \propto a_\mathrm{V}^{-3.5} \mathrm{d}a_\mathrm{V}.
\end{equation}
Using $a_\mathrm{V}$ provides a common size variable for compact spheres, irregular grains, and aggregate particles. It is the convention adopted in the {\tt{AggScatVIR}} database to compute size-distribution-averaged optical properties \citep{tazaki_fractal_2023}. For all models, we fixed the lower limit of the distribution to $a_{\mathrm{V},\mathrm{min}} = 0.2~\mu$m, corresponding to the smallest particle size available in {\tt{AggScatVIR}}. The upper limit $a_{\mathrm{V, max}}$ ranges from $0.2$ to $1.6~\mu$m for irregular grains and up to $\sim2~\mu$m for the aggregates models. 
However, the volume-equivalent radius $a_\mathrm{V}$ does not represent the apparent size of aggregates, so we used the characteristic radius, denoted $a_\mathrm{c}$ \citep{mukai_radiation_1992, kozasa_optical_1992} and determined from the radius of gyration $a_g$. It quantifies the spatial distribution of monomers around an aggregate's center of mass. The characteristic radius $a_\mathrm{c}$ is better suited to comparing the overall dimensions of aggregates with different morphologies. We denote with $a_{c,\mathrm{max}}$ the characteristic radius of the largest aggregate included in a given model. Depending on the aggregate morphology and monomer size, $a_{c,\mathrm{max}}$ reaches values of up to $\sim 12~\mu$m.

\subsection{Complex dust particle in MCFOST} \label{Sect:Mueller}
The {\tt{MCFOST}} 3D radiative transfer code can now model light scattering by nonspherical grains, such as dust aggregates, through the full Stokes–Mueller formalism ($I,~Q,~U,~V$) during the propagation of Monte Carlo photon packets. To achieve this, the scattering (Mueller) matrix is directly imported into {\tt{MCFOST}}. Its elements ($S_{11}$, $S_{12}$, $S_{22}$, $S_{33}$, $S_{34}$, and $S_{44}$) are taken from the {\tt{AggScatVIR}} database \citep{tazaki_fractal_2023} following the definition of \citet{bohren_absorption_1983} as a function of scattering angle and wavelength. These scattering matrix elements and the corresponding extinction and absorption opacities are averaged over the grain size distribution (see Sect. \ref{Sect:Dustmodel}). All $S_{ij}$ elements are initially expressed in units of $\mathrm{sr}^{-1}$, but for radiative transfer calculations with {\tt{MCFOST}}, the scattering matrix is then scaled such that only $S_{11}$ holds the unit $\mathrm{sr}^{-1}$, while the other components are used as dimensionless values $S_{ij}/S_{11}$. This new feature of {\tt{MCFOST}} enables the generation of synthetic images of the scattered light and polarization of protoplanetary disks composed of complex dust models such as aggregates.

\section{Results on disk colors} \label{Sect:Results}
\subsection{Measuring the color from disk observations} \label{Sect:Measuring}
\begin{figure}
    \centering
    \includegraphics[width=0.99\linewidth]{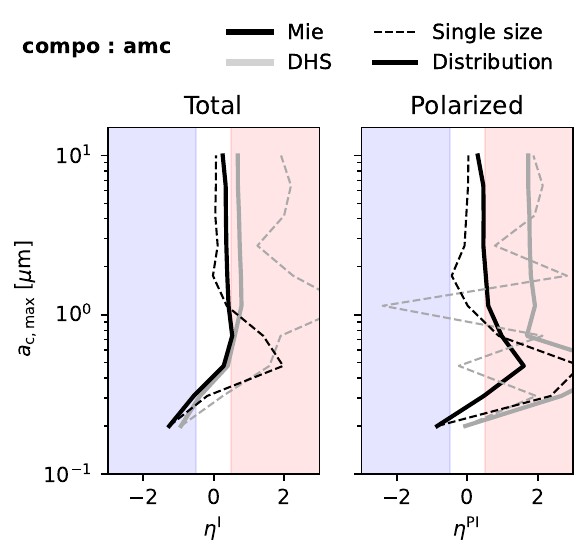}
    \caption{Disk color $\eta_\mathrm{J/H}$ with respect to the maximum grain size for the Mie (black) and DHS (gray) dust models using an amc composition for both the total (\textit{left}) and polarized (\textit{right}) intensity. We distinguish between a size distribution and single size using solid and dashed curves, respectively. Blue and red respectively correspond to $\eta<-0.5$ and $\eta>0.5$ disk color.}
    \label{fig:Mie_color}
\end{figure}
\begin{figure*}
    \centering
    \includegraphics[width=0.99\linewidth]{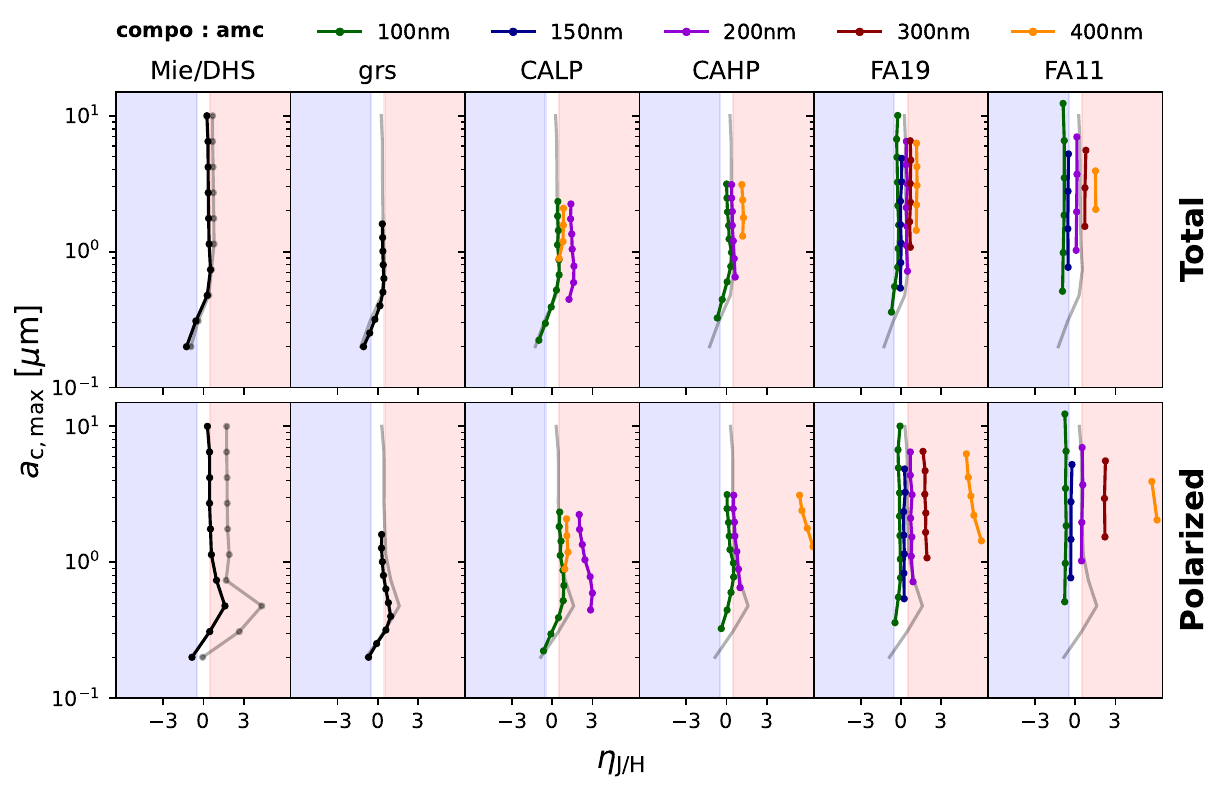}
    \caption{Panel of $\eta_{J/H}$ for total (\textit{top}) and polarized (\textit{bottom}) intensity for grains model with an amc composition. From left to right, $\eta_{J/H}$ is displayed for spherical grains (Mie theory in black and DHS in gray), GRS, CALP, CAHP, FA19, and FA11. For a better comparison, the case of the Mie dust model is plotted on all graphs as gray curves. For aggregates, we distinguish measurements by monomer size: $100$nm (green), $150$nm (blue), $200$nm (purple), $300$nm (red), and $400$nm (orange).}
    \label{fig:PanelColor}
\end{figure*}

The total intensity disk color ($\eta$) is defined by the gradient of scattered light between two observational wavelengths, $\lambda_1$ and $\lambda_2$ \citep{tazaki_effect_2019},
\begin{equation}\label{EQ:color}
    \eta^{I}_{\lambda_1, \lambda_2} = \frac{\log(I_d/I_\star)_{\lambda_2} - \log(I_d/I_\star)_{\lambda_1}}{\log(\lambda_2/\lambda_1)},
\end{equation}
where $I_\star$ is the  total stellar intensity, and $I_d$ is the integrated total scattered intensity coming only from the outer disk surface, and can be substituted by the polarized intensity $Q_\phi$, to obtain the polarized disk color $\eta^{PI}$. 
For some dust models, especially when there is no size distribution (single size), the integrated polarized intensity can be negative, resulting in discontinuities in the disk color computation. In such cases, we used the absolute value of $Q_\phi$ in the color computation. We defined three ranges of disk colors using this measurement: blue for $\eta<-0.5$, gray when $-0.5<\eta<0.5$, and red if $\eta>0.5$ \citep{tazaki_effect_2019, ma_color_2024}. 

The color of the scattered light depends mainly on the dust properties and not on the disk geometry, such as inclination and scale height, because the scattering geometry is expected to be very similar at different wavelengths between the visible and near-infrared \citep{ma_color_2024}. We verified the validity of this assumption in our models by testing the impact of inclination and vertical structure on the measured disk colors in Appendix~\ref{APP:Geometry}, and the presence of an inner disk in Appendix ~\ref{APP:InnerDisk}.

To investigate the impact of dust properties on disk color, we computed the disk color for all the available dust models and group them by particle type. We then represented this color as a function of the characteristic radius of the grain $a_\mathrm{c}$. For aggregate models, we further distinguished between the different monomer sizes to better illustrate their impact on the disk color.

In the following, we focus on the near-infrared disk color between  the J band (centered at $1.25 \mu$m) and the H band (centered at $1.63 \mu$m) denoted as $\eta_\mathrm{J/H}$. This choice is motivated by the large number of scattered-light observations available in these two bands, in particular from VLT/SPHERE \citep{beuzit_sphere_2019, ginski_sphere_2024, ginski_disk_2025, roumesy2026}. Figure~\ref{fig:Mie_color} illustrates the color measurement for the Mie and DHS models, comparing the results obtained for a single grain size with those obtained for a grain size distribution. Single size dust models show strong variations in color with $a_\mathrm{c}$, especially in polarized intensity. These variations arise from the rapid changes in scattering and polarization efficiencies with the grain size due to the Mie scattering. As the size distribution becomes broader, the contributions from grains of different sizes average out the Mie resonances, resulting in smoother color trends. As protoplanetary disks are expected to contain grains with sizes distributed over a relatively wide range, we adopted the grain size distribution defined in Sect. \ref{Sect:Dustmodel} for the rest of this study. Figure \ref{fig:PanelColor} and the following sections present the results obtained for the representative subset of dust models introduced in Sect. \ref{Sect:Dustmodel} (Mie/DHS, GRS, CALP, CAHP, FA19, and FA11). The complete set of color results for the two dust compositions (org and amc), as well as the colors obtained considering other wavelengths, are presented in Appendix~\ref{APP:AllColor}.

\subsection{Color trends for compact and irregular grains} \label{Sect:Compact}
Figure \ref{fig:PanelColor} summarizes the color results from model calculations for both total intensity (\etaI) and polarized intensity (\etaPI). For spherical grains, both \etaI~and \etaPI\  are  steady at the limit between gray and red for large grains. In total intensity, the color is blue for $a_\mathrm{c,max} \leq 1~\mu$m, whereas polarized intensity shows a bump from blue to red at $a_\mathrm{c,max} \sim 0.5~\mu$m. The total intensity color \etaI~is similar for Mie and DHS particles. However, the DoLP is largely affected by the shape of the particles, explaining why the \etaPI~is redder for DHS. We also observe a bump toward the red in polarized intensity at $a_\mathrm{c, max} \sim0.5~\mu$m. 

For GRS, the total and polarized intensity colors follow the same trend as for spherical particles based on Mie theory. We note, however, that the polarized intensity color can deviate from the Mie approximation depending on the composition (see Appendix \ref{APP:AllColor}). The Mie theory still provides a good approximation for GRS, at least for total intensity.

\subsection{Color trends for aggregates} \label{Sect:Aggregates}
For more complex particles like compact aggregates, we still have a general trend similar to spherical grains (Mie), but significant deviations start to arise, especially for CAHPs with higher porosity. Already visible for CALPs, the bump toward red at $\sim 0.5 \mu$m observed on \etaPI~is less pronounced than in the case of Mie theory. In addition, the monomer size seems to have a stronger impact on the polarized color than on the total intensity color: the DoLP is affected by particle shape, so larger monomers tend to produce a redder disk polarized color.

In the case of a disk composed of fractal aggregates, the bump toward red polarized color observed for spherical grains completely disappears. The key finding here is that both colors \etaI~and \etaPI~are independent of the overall particle size ($a_\mathrm{c}$). However, they clearly show a strong sensitivity to the size of the monomers that compose the aggregate. This behavior was suggested by SPHERE observations in \cite{roumesy2026} and highlights that the chromatic properties of light scattered by fractal aggregates are primarily governed by the substructure of the grains rather than their overall dimensions. We detail this in Sect. \ref{Sect:Monomers}. 

These results show that a blue disk color is not only associated with small spherical grains, but can also be reproduced by fractal aggregates composed of small monomers, even when the overall aggregate size is larger than the observing wavelength. Fractal aggregates composed of larger monomers tend to produce redder colors, illustrating that the color of a disk containing aggregates is primarily sensitive to the substructure of the particle rather than to its overall size. For smaller monomers resulting in aggregates much smaller than the diffusion scale, $a_\mathrm{c} \ll \lambda/(2\pi)$, the scattering is expected to progressively approach the Rayleigh regime and the disk color should become bluer. For the H band considered here ($1.63~\mu$m), this characteristic scale is $\lambda/(2\pi) \simeq 0.25~\mu$m.

From a more general point of view, and comparing the disk polarized color produced by spherical dust grains with other dust models, we observe that the dependence on grain size differs from what is found for spherical grains when the dust shape becomes more complex. In other words, the shift toward the reddish polarized color at $\sim0.5\mu$m, clearly visible for spherical grains, attenuates and then disappears as shape complexity increases (from left to right in Fig. \ref{fig:PanelColor}).

\subsection{The key role of monomer size} \label{Sect:Monomers}
\begin{figure}
    \centering
    \includegraphics[width=0.99\linewidth]{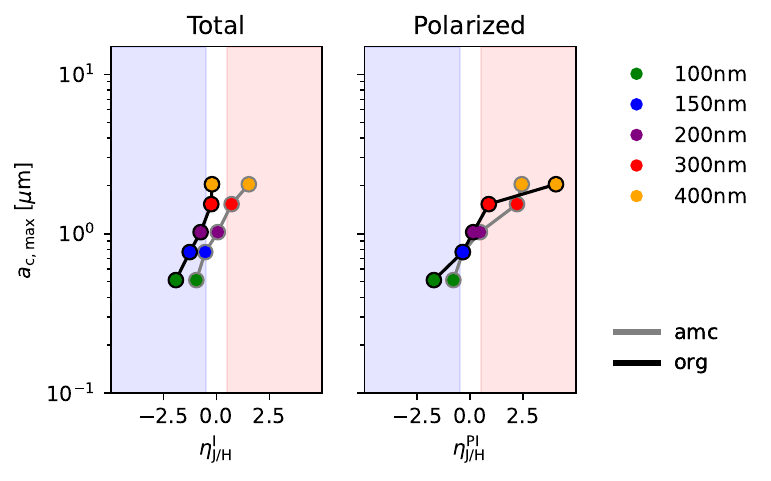}
    \caption{Impact of monomer size on near-infrared disk color observed in total (\textit{left}) and polarized (\textit{right}) intensity. We show the case of the FA11 dust model, with org (black) and amc (gray) compositions.}
    \label{fig:monomers}
\end{figure}

In the case of dust aggregates, we find that the particle size does not significantly affect color, whereas the size of the monomers plays a key role for both \etaI~and \etaPI. To emphasize this result, the dependence of the near-infrared color $\eta_{J/H}$ on monomer size is shown in Fig. \ref{fig:monomers}, keeping the same number of monomers for a fractal aggregate with dimension $D_f = 1.1$ (FA11 dust model). 

We find that the disk color becomes redder as the monomer size increases for fractal aggregates, especially for polarized color \etaPI. This trend is clear for fractal models, but is not systematically recovered for compact aggregates. For example, the CALP model does not exhibit this increasing trend: aggregates composed of $200$~nm monomers produce redder colors than those composed of $400$~nm monomers. This dependence of disk color varies with aggregate porosity, and the CAHP model returns to a trend closer to the case of fractal aggregates, with redder colors for larger monomers. These trends remain qualitatively similar independently of the number of monomers considered.

Nevertheless, and together with the fact that the color remains constant for a given monomer size, a key result of this study is that the disk color is governed by the size of the monomers that constitute the aggregate. However, the color remains different from the color found on a disk composed of monomer-sized spherical particles. This suggests that the structure of aggregates also plays a role in determining the disk color.

The insensitivity of the disk color to aggregate size is a natural consequence of the interference of scattered waves. For low-dimension fractal aggregates, multiple scattering is usually negligible, and the scattered wave can be described as the superposition of waves scattered by individual monomers \citep{tazaki_light_2016}. Once the aggregate size becomes larger than the wavelength, waves scattered by monomers that are widely separated within the aggregate become out of phase, resulting in incoherent scattering. The scattering opacity is therefore primarily dominated by coherent scattering from monomers located within a region comparable to the wavelength scale (see also Fig. 7 in \citealp{tazaki_effect_2019}). Ultimately, the scattering properties are determined by the number density and spatial arrangement of monomers on the wavelength scale, rather than by the overall size of the aggregate. Furthermore, since it is a superposition of scattered waves from each monomer, it strongly reflects the properties of monomers, leading to its key role in determining the color.

\section{Discussion} \label{Sect:Discussion}
\subsection{Comparison with simple light scattering models} \label{Sect:EMT}
In radiative transfer calculations of protoplanetary disks, the effect of porosity is often considered using the effective medium theory (EMT) with Mie theory or DHS (e.g., \citealt{min_modeling_2005, min_optical_2009}). Here, we discuss the extent to which these simple approaches reproduce the trends found in Sect. \ref{Sect:Results}, which are based on detailed light scattering physics. 

In the EMT framework, an aggregate is replaced by a homogeneous spherical grain with the same size and porosity\footnote{A proper choice of the radius of the effective-medium spherical particle is not trivial. One common approach is to use the characteristic radius $a_\mathrm{c}$ \citep{mukai_radiation_1992, kozasa_optical_1992}, which we adopt here.} and we adopted the Bruggeman mixing rule \citep{Bruggeman1935}. We first used it to examine how porosity alone affects the disk color, and then tested whether it can reproduce the colors obtained with the exact aggregate optical properties.

\begin{figure}
    \centering
    \includegraphics[width=0.99\linewidth]{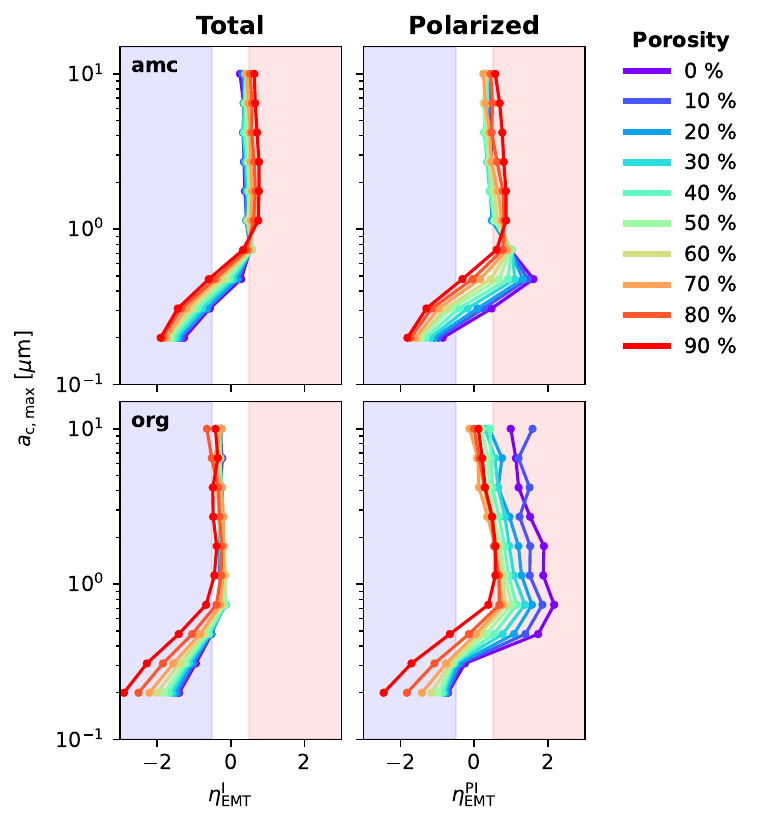}
    \caption{Effect of the dust porosity on the disk color $\eta_\mathrm{J/H}$ for total (\textit{left}) and polarized (\textit{right}) intensity, considering both amc (\textit{top}) and org (\textit{bottom}) compositions. The disk is modeled using a size distribution of spherical grains described by the EMT approach.}
    \label{fig:porosity}
\end{figure}

Figure \ref{fig:porosity} first shows the effect of porosity on the disk color $\eta_\mathrm{J/H}$: For the org composition, the polarized intensity disk color is much redder for lower porosity, and shifts toward bluer colors when the dust porosity increases. For more absorbing material (amc), compact grains produce a bump toward the red in \etaPI~at characteristic radii around $a_{\mathrm{c}} \simeq 0.5\mu$m. This feature progressively weakens and eventually disappears as the porosity increases. Both the total and the polarized intensity colors shift toward the blue with increasing porosity for $a_{\mathrm{c}} \lesssim 0.9~\mu$m. The opposite trend is observed for larger grains, although the effect of porosity is weaker. Ultimately, porosity has a significant impact on the disk color and seems to be more pronounced in polarized intensity than in total intensity. Note that its impact on color is influenced by the dust grain composition.

To test whether the EMT can reproduce the disk colors obtained with the exact aggregate optical properties, we generated synthetic disk images composed of spherical dust grains model, with the same characteristic radius and porosity as the aggregate models from our representative subset (i.e., CALP, CAHP, FA19, and FA11) considering the same power-law size distribution. For instance, the values for the characteristic radius and porosity of compact and fractal aggregates are directly imported from the {\tt{AggScatVIR}} database for each size (i.e., each number of monomers) and used in the {\tt{MCFOST}} simulations. In the EMT approach, the Bruggeman and Maxwell--Garnett mixing rules assume that the monomers are sufficiently small compared with the wavelength. Aggregates composed of larger monomers start to violate this condition. We therefore focus here on a monomer size of $100$ nm. In the following, we denote with $\eta_\mathrm{EMT}$ the color measured on these EMT-based disks, and by $\eta_\mathrm{agg}$ the color obtained in Sect.~\ref{Sect:Results} using the exact optical properties of the aggregates computed with the T-matrix method. The aim is to quickly determine which dust properties are well approximated by the EMT method, knowing that the agreement between $\eta_\mathrm{EMT}$ and $\eta_\mathrm{agg}$ depends on both the aggregate porosity and the monomer size. 

\begin{figure}
    \centering
    \includegraphics[width=0.99\linewidth]{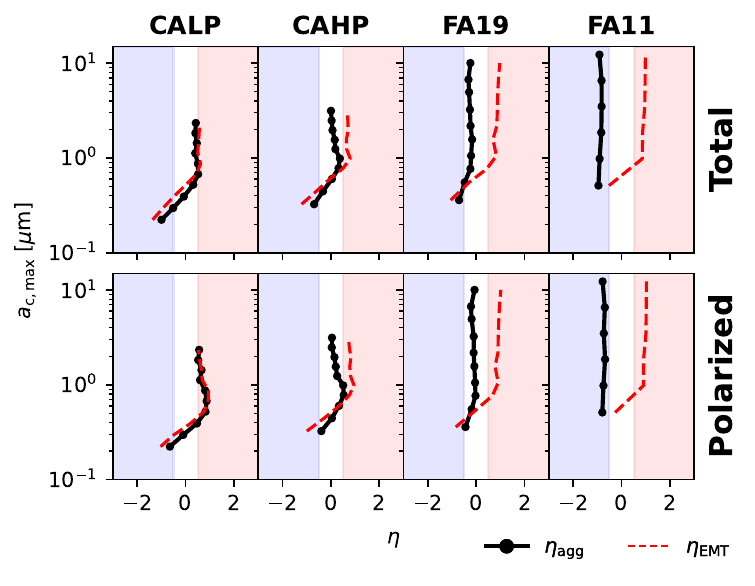}
    \caption{Comparison between the disk colors $\eta_\mathrm{J/H}$ obtained with the exact aggregate optical properties ($\eta_\mathrm{agg}$) in black and with the EMT method ($\eta_\mathrm{EMT}$) in dashed red. Both models consider aggregates with monomer radii of $100$ nm and amc compositions. The top and bottom rows correspond to total and polarized intensity, respectively and aggregate models are ordered according to increasing porosity from left to right.}
    \label{fig:EMT}
\end{figure}

Figure \ref{fig:EMT} shows that the EMT approximation reproduces the color for compact aggregates with lower-porosity (CALP) and deviations start to occur for higher-porosity aggregates (toward the FA11 dust model). This trend is observed in both total and polarized intensity. Surprisingly, the EMT predicts the opposite trend in color with porosity compared to $\eta_\mathrm{agg}$: the disk becomes redder as porosity increases, whereas true calculations (based on the exact optical properties of the aggregate) indicate a shift toward bluer colors. However, this opposite trend is also related to the composition. 

Previous studies have shown that, although EMT can provide a first-order, and in some cases reasonable, approximation to the optical properties of less porous dust aggregates or pseudo-porous spheres \citep[e.g.,][]{Voshchinnikov2005, Voshchinnikov2007, shen_modeling_2008, shen_modeling_2009, Draine2025}, it fails to reproduce the optical properties of dust aggregates when they have a fractal dimension lower than three (such as the FA19, FA11 models; \citealt{tazaki_light_2016, tazaki_light_2018}) and/or when the size of the inclusions (i.e., the monomer size) becomes comparable to or larger than the wavelength \citep[e.g.,][]{Chylek2000}. Our results extend this conclusion to the colors of protoplanetary disks. Although the EMT method can reproduce the disk colors for smaller monomers and relatively compact aggregates, it becomes unreliable for aggregates with large monomers and high porosity. Consequently, disk colors derived from EMT model $\eta_\mathrm{EMT}$ should be interpreted with caution when the particles are aggregates whose monomer size is not negligible compared with the observing wavelength.

\subsection{Can the albedo predict disk colors?} \label{Sect:Albedo}
In previous work, the albedo was often used to evaluate the influence of dust properties on the scattered light of protoplanetary disks. 
In this section we compare predictions based on effective albedo with our measurements of the color. The aim is to determine to what extent the effective albedo can be used to predict the color of a disk without performing time-consuming radiative transfer calculations.

\begin{figure}
    \centering
    \includegraphics[width=0.99\linewidth]{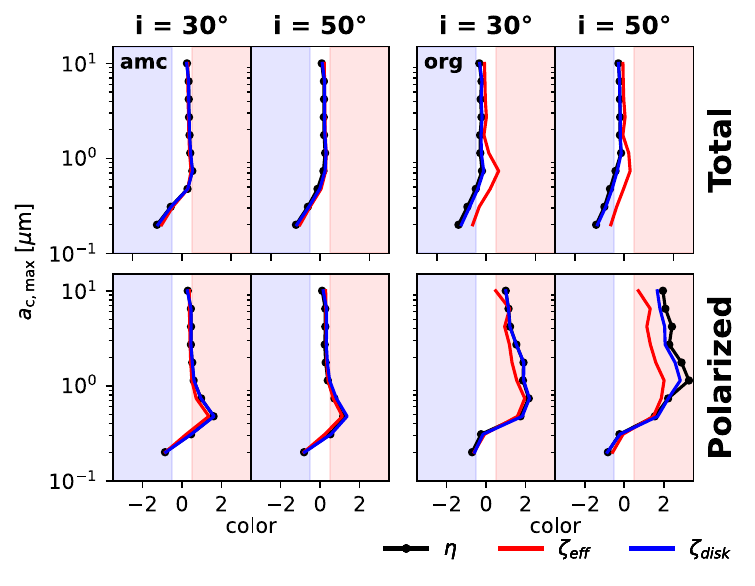}
    \caption{Comparison between the disk color $\eta_{J/H}$ and logarithmic gradient based on the effective albedo (\zetaeff) and the disk albedo (\zetadisk) for two dust compositions -- amc (\textit{left}) and org (\textit{right}) -- in total and polarized intensity. For each composition, the results are presented for two disk inclinations: $i = 30^\circ$ and $i = 50^\circ$. The measured color ($\eta$) is plotted in black, \zetaeff~in red, and \zetadisk~in blue.}
    \label{fig:color_vs_albedo}
\end{figure}
The dust albedo $\omega$ quantifies the fraction of light scattered by a dust grain, in contrast to extinction. In a protoplanetary disk image, however, only a limited range of scattering angles contributes to the observed scattered light. The 
disk inclination and scattering surface geometry set the range of accessible scattering angles. To account for this limited scattering angle range, we used the effective albedo, denoted as $\omega_{\mathrm{eff}}$, following \citet{mulders_why_2013}. It was computed using the scattering phase function $S_{i, j}(\theta)$ between the accessible scattering angles $\theta_1$ and $\theta_2$, normalized by the integrated total intensity phase function $S_{11}$:
\begin{equation} \label{EQ:effective_albedo}
    \omega_{\mathrm{eff}} = \frac{2\omega\int_{\theta_1}^{\theta_2}S_{i,j}(\theta)\sin\theta d\theta}{ (\cos\theta_1 - \cos\theta_2) \int_{0}^{\pi}S_{11}(\theta)\sin\theta d\theta}.
\end{equation}
Here, we define $S_{i, j}(\theta) = S_{11}$ for total intensity and $S_{i, j}(\theta) = S_{12}$ for polarized intensity. An effective albedo that increases toward longer wavelengths is expected to produce a redder disk color, while a decreasing effective albedo is expected to produce a bluer one.

For a consistent comparison between the wavelength dependence of $\omega_\mathrm{eff}$ with the disk color, we define $\zeta^{\mathrm{eff}}_{\lambda_1,\lambda_2}$, the logarithmic gradient of effective albedo between two wavelengths $\lambda_1 < \lambda_2$. We use the same definition as for the measured color (see Eq. \ref{EQ:color}):
\begin{equation}
    \zeta^{\mathrm{eff}}_{\lambda_1,\lambda_2} = \frac{\log(\omega_{\mathrm{eff}}(\lambda_2)) - \log(\omega_{\mathrm{eff}}(\lambda_1))}{\log(\lambda_2/\lambda_1)}.
\end{equation}

Figure \ref{fig:color_vs_albedo} compares the color ($\eta$) measured from the images produced with radiative transfer with the predicted color based on effective albedo \zetaeff~using the optical properties of the dust. For these tests, we considered spherical particles according to Mie theory, with two different compositions (org and amc). As the effective albedo is determined by the range of accessible scattering angles, we also examined two disk inclinations ($30^\circ$ and $50^\circ$).

For an amc composition, the effective albedo successfully reproduces the measured color in both total and polarized intensity. However, discrepancies occur for the org composition, where the effective albedo predicts a redder color than that measured in the disk images, in total intensity (\etaI). The trend is reversed in polarized intensity, where the effective albedo estimates a bluer color than \etaPI, the color measured from the synthetic images. These general findings also apply to aggregates (not shown here).

Regarding the dust composition, the higher reflectivity of the org composition may increase the contribution of multiple scattering, which is not captured by the single-scattering approximation underlying \zetaeff. In polarized intensity, this effect is more significant at forward scattering angles \citep{tazaki_fractal_2023}. Scattering angle range increases with the inclination of the disk, resulting in more forward scattering. We suggest that this higher sensitivity to multiple scattering  explains the discrepancy between \zetaeff~and $\eta$ at higher inclinations.

In summary, using effective albedo as an indicator of disk color appears to be a first approach for reducing the range of dust parameters to be examined in the study of disk color. However, an in-depth analysis using radiative transfer simulations is still necessary to ensure a consistent interpretation of the colors and dust properties within the disks.

\subsection{The case of HD 100546 total intensity color} \label{Sect:HD100546}
After studying the influence of dust properties on the color of protoplanetary disks, we applied our results to a specific disk. The aim was to determine whether combining multiwavelength observations can better constrain the estimation of dust properties and break the degeneracy regarding which dust models can reproduce the observed scattered-light measurement \citep{roumesy2026}. We consider the example of the disk around HD~100546, which was observed in scattered light by the \textit{Hubble} Space Telescope (HST) across the optical to near-infrared wavelengths, using ACS (Advanced Camera for Surveys) with F435W and F814W filters \citep{Ardila2007}, and HST/NICMOS with F160W \citep{Augereau2001} and F222M filters (part of Program GO 10167 PI:Weinberger). Based on these observations, \citet{mulders_why_2013} revealed a strong red color from the disk, which they attributed to relatively large dust grains (between $2.5$ and $10~\mu$m), assumed to be porous due to their presence on the disk surface. This reddish color was estimated using the disk albedo, measured directly from the images, similar to the color measurements presented here.

The disk albedo, denoted as $\omega_\mathrm{disk}$, is derived from the surface brightness of the disk $\mathrm{SB}(r)$ as defined in \cite{mulders_why_2013}. They define the fraction of light scattered toward the observer as the geometric albedo $\omega_\mathrm{geo} = \mathrm{SB}(r)/((F_\star+F_\mathrm{inner})/4\pi r^2)$ where $F_\star$ and $F_\mathrm{inner}$ are the flux densities from the star and from the inner disk ($r\leq 1$~au), respectively. The disk albedo $\omega_\mathrm{disk}$ is then obtained by modeling the radial dependence of $\omega_\mathrm{geo}$ using the power law $\omega_\mathrm{geo}(r) = \omega_\mathrm{disk} \cdot r^{-q}$, where $q$ represents the disk outward attenuation. 

Similar to our approach for the effective albedo, we compare the logarithmic gradient of the disk albedo \zetadisk~measured from the models to evaluate its agreement with the measured color $\eta$. Figure \ref{fig:color_vs_albedo} reveals that \zetadisk~clearly follows $\eta$ in both total and polarized intensity, as expected since the disk albedo and color are measured directly from the disk image. A slight discrepancy can arise when the disk inclination is larger, due to the disk geometry: projection effects and integration along the line of sight impede the measurement of the surface brightness $\mathrm{SB}(r)$ at different disk radii. 

Therefore, we can compare the disk albedo values measured on HD~100546 by \cite{mulders_why_2013} with the colors obtained through our models. These measurements of $\omega_\mathrm{disk}$ are not obtained at the same wavelengths as our models. To ensure a consistent comparison, we linearly interpolate the measurement to our model wavelengths of $0.554$, $0.735$, and $1.63~\mu$m. These wavelengths are selected based on their proximity to the observed ones. We can now measure the estimated colors on HD~100546 $\zeta_\mathrm{disk} (0.554~\mu \mathrm{m},~ 0.735~\mu \mathrm{m}) = 0.612$ and $\zeta_\mathrm{disk} (0.735~\mu m,~1.63~\mu m) = 1.338$ to directly compare them with our results presented in Sect. \ref{Sect:Results}. The values of $\omega_\mathrm{disk}$ reported by \citet{mulders_why_2013} are given without uncertainties, so the errors on \zetadisk~cannot be propagated. We therefore adopt a conservative uncertainty of $0.25$ for both measurement of \zetadisk. Note that our model assumes an inclination angle of 30$^\circ$, which is smaller than that of the HD~100546 disk (42$^\circ$; \citealp{Ardila2007}). However, the inclination has a negligible impact on the disk colors and therefore does not affect our conclusions (see Appendix \ref{APP:Geometry}).

Figure \ref{fig:HD100546Color} shows the color-color diagram of disk scattered light for the models and the estimation from HD 100546. First, compact grains (Mie, DHS, or GRS) fail to reproduce the red colors observed. Similarly, models with org composition mostly produce blue or gray colors, which do not match the observations. However, a key finding stands out: some aggregate models — fractals ($a_\mathrm{mon} = 400$~nm), CALPs ($a_\mathrm{mon} = 200$~nm), and CAHPs ($a_\mathrm{mon} = 400$~nm) — are remarkably consistent with the measurements on HD 100546 when they are composed of more absorbing material (amc). Furthermore, the colors for the different dust models remain clearly distinguishable even when a conservative uncertainty of $0.25$ is applied to \zetadisk. 
This example based on HD~100546 illustrates the advantage of performing a multiwavelength analysis to better constrain the dust properties. These results lend further support to the importance of considering the presence of aggregates in protoplanetary disks.
\begin{figure}
    \centering
    \includegraphics[width=0.99\linewidth]{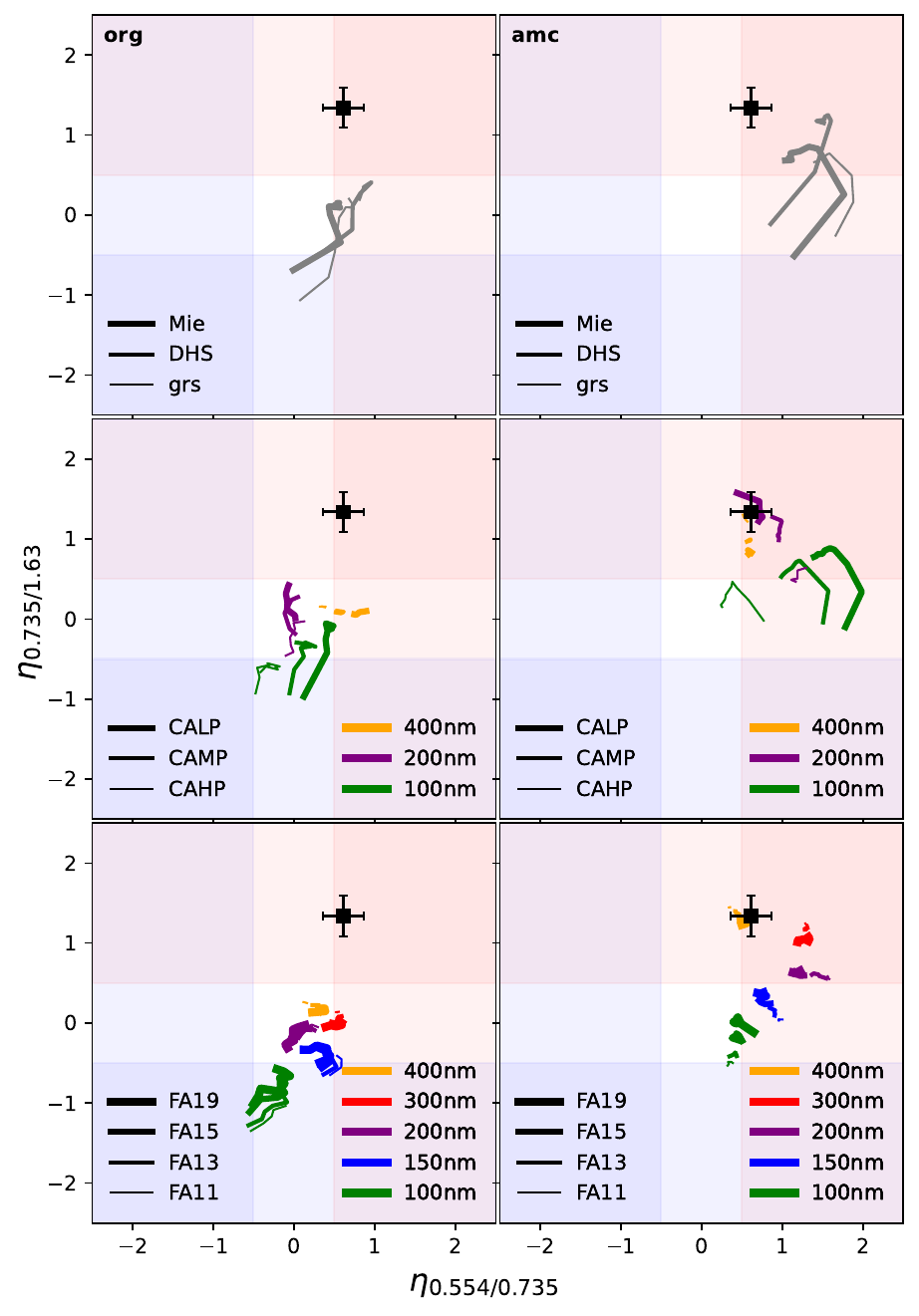}
    \caption{Color-color diagram between $\eta_{0.554/0.735}$ and $\eta_{0.735/1.63}$ in total intensity for all the dust models defined in Sect. \ref{Sect:Dustmodel}. We separate the plots with respect to the dust composition -- org (\textit{left column}) and amc (\textit{right}) -- and with respect to the dust family (from top to bottom): compact grains (Mie, DHS, GRS), compact aggregates (CALP, CAMP, CAHP), and fractal aggregates (FA19, FA15, FA13, FA11). We also plot the color estimated on HD 100546 \citep{mulders_why_2013} as a black cross, with uncertainty arbitrarily fixed to 0.25. For aggregates, the models are separated with respect to the monomer size. Blue and red correspond to $\eta<-0.5$ and $\eta>0.5,$ respectively, for both disk colors (vertically and horizontally).}
    \label{fig:HD100546Color}
\end{figure}

Interestingly, recent \textit{James Webb} Space Telescope observations of edge-on disks around HH~30 \citep{tazaki_jwst_2025}, Tau 042021 \citep{duchene_jwst_2024}, Oph~163131, and IRAS~04302+2247 \citep{villenave_jwst_2024, villenave_jwst_2024-1} suggest the presence of very large grains (between $3$ and $10\mu$m), which are often considered to be fully vertically mixed to explain their presence in the disk surface layers. As in the case of HD~100546, the presence of these large dust grains on the surface of the disk is challenging to justify with turbulence and vertical settling. Following this analysis, we can speculate on the presence of large porous aggregates within these edge-on disks, which may be coupled to the gas and remain on the surface of the disk. 

\subsection{Caveats} \label{Sect:Caveats}
Regarding the colors measured on the disks, the disk inclination and surface height only slightly affect the measured colors for face-on to moderately inclined disks (see Appendix \ref{APP:Geometry}). These geometric parameters do not alter the main findings of this study. However, for a targeted analysis of a single disk, it would be beneficial to adjust the geometry of the synthetic disks derived from radiative transfer, particularly their inclination, to better match the observed disk. This geometric consistency enables a more robust comparison between observations and simulations and allows for a better interpretation of dust properties based on the colors.

One major assumption of the {\tt{AggScatVIR}} aggregate dust models is that dust aggregates are monodisperse, meaning that all the monomers forming an aggregate are assumed to have a single size. 
In this study, we have shown that disk color is mainly governed by monomer size rather than by the overall size of the particle. Therefore, incorporating a polydisperse distribution of monomer sizes could have a significant impact on the colors observed in scattered light. In future studies, improving our understanding of the dust properties in protoplanetary disks will therefore require simulations of scattering that incorporate polydisperse monomers.

\section{Conclusion} \label{Sect:Conclusion}
In this work, we have investigated the impact of complex dust particles on the color of protoplanetary disks in both total and polarized intensity. Using the radiative transfer tool {\tt{MCFOST}} together with the accurate scattering properties of irregular, compact aggregate, and fractal aggregate particles from the {\tt{AggScatVIR}} database, we explored disk colors beyond the simplifying assumptions of classical Mie theory.

\begin{itemize}
    \item We have found that the dependence of disk color on the overall particle size becomes weaker as particle morphology becomes more complex. While compact spherical grains show a pronounced variation of polarized color with grain size, this feature weakens for irregular and porous particles and vanishes for fractal aggregates. Both total and polarized intensity colors are largely independent of the aggregate characteristic radius and are instead primarily governed by the monomer size. Larger monomers generally produce redder colors, whereas small monomers produce blue colors. Consequently, a blue color does not necessarily imply the presence of small individual grains: it can also be produced by large, fluffy aggregates composed of small monomers. Dust composition also plays a key role: the more absorbing amc composition produces a wider range of colors, from blue to strongly red colors, whereas the organic-rich composition (org) generally leads to blue or neutral colors.
    \item Our simulations also show that polarized intensity is a more sensitive indicator of the morphology and size of monomers than total intensity. The polarized color therefore provides a better diagnostic for interpreting the properties of dust in protoplanetary disks. This result strengthens the need to develop polarimeters for future large telescopes, as systematic polarized observations at multiple wavelengths could provide valuable constraints on the properties and evolution of dust in the scattering surface of protoplanetary disks. 
    \item We show that the EMT provides a reasonable approximation for relatively compact aggregates made of small monomers but fails to reproduce the colors of highly porous aggregates or aggregates composed of larger monomers.
    \item As an additional observation, we have shown that the effective albedo accurately reflects the disk color trends and can be used to narrow down the range of dust models with amc composition. However, the effect of multiple scattering increases for more reflective particles (org composition), reducing the reliability of color predicted with the effective albedo. To overcome this limitation and account for multiple scattering, the use of radiative transfer simulations is necessary to perform a quantitative interpretation of the disk color. 
    \item The color-color analysis of HD~100546 illustrates the diagnostic power of multiwavelength measurements. Its strongly red optical to near-infrared colors are not reproduced by compact-particle models or by the org composition, but are consistent with several aggregate models composed of absorbing material (amc) and relatively large monomers.

\end{itemize}

Overall, disk color provides a valuable estimation of dust properties in protoplanetary disks, with a particular sensitivity to the monomer size in aggregate particles. Its diagnostic power can be strengthened by combining multiple color measurements together or with polarized scattering phase functions. Applying this framework to individual disks, supported by the growing number of multiwavelength polarized observations of disks, will help break degeneracies between dust size, composition, morphology, and aggregate substructure. Such combined analyses will provide stronger constraints on dust evolution through coagulation, fragmentation, settling, and radial drift, thereby improving our understanding of the physical processes that lead to planet formation.

\begin{acknowledgements}
This project has received funding from the European Research Council (ERC) under the European Union Horizon Europe research and innovation program (grant agreement No. 101053020, project Dust2Planets, PI: F. M\'enard). The authors thank the IDEX for funding MR mobility grant. Special acknowledgement must also go to F. Malaval for his work on developing the MCFOST option to implement complex particles in the radiative transfer calculation. R.T. was supported by JSPS KAKENHI Grant No. JP25K07351.
\end{acknowledgements}

\bibliographystyle{aa}
\bibliography{Biblio}

\onecolumn
\appendix
\nolinenumbers

\section{Dust models} \label{APP:DustModels}
This section presents a table of all the dust models used in our study (see Table \ref{tab:dustmodels}). We applied Mie theory for compact spherical particles and the DHS model to approximate the scattering properties of aggregates. We also used models from the {\tt{AggScatVIR}} database, which includes scattering properties for particles with much more complex shapes.

\begin{table*}[h!]
    \centering
    \caption{Dust models parameters used in simulations of disk images.}
    {\fontsize{7.2}{10}\selectfont
    \begin{tabular}{ c c c c c c}
    \hline
    \bottomrule
        model & 100nm & 150nm & 200nm & 300nm & 400nm \\
        \hline
        FA11 & 8, 16, 32, 64, 128, 256                        & 8, 16, 32, 64                   & 8, 16, 32, 64                     & 8, 16, 32                 & 8, 16 \\
        FA13 & 8, 16, 32, 64, 128, 256, 512                   & 8, 16, 32, 64, 128              & 8, 16, 32, 64, 128                & 8, 16, 32, 64             & 8, 16, 32 \\
        FA15 & 8, 16, 32, 64, 128, 256, 512, 1024             & 8, 16, 32, 64, 128, 256         & 8, 16, 32, 64, 128, 256           & 8, 16, 32, 64, 128        & 8, 16, 32, 64 \\
        FA19 & 8, 16, 32, 64, 128, 256, 512, 1024, 2048, 4096 & 8, 16, 32, 64, 128, 256, 512    & 8, 16, 32, 64, 128, 256, 512      & 8, 16, 32, 64, 128, 256   & 8, 16, 32, 64, 128 \\
        CAHP & 8, 16, 32, 64, 128, 256, 512, 1024, 2048, 4096 & ...                             & 8, 16, 32, 64, 128, 256, 512      & ...                       & 8, 16, 32, 64 \\
        CAMP & 8, 16, 32, 64, 128, 256, 512, 1024, 2048, 4096 & ...                             & 8, 16, 32, 64, 128, 256, 512      & ...                       & 8, 16, 32, 64 \\
        CALP & 8, 16, 32, 64, 128, 256, 512, 1024, 2048, 4096 & ...                             & 8, 16, 32, 64, 128, 256, 512      & ...                       & 8, 16, 32, 64 \\
        \midrule
        GRS  & \multicolumn{3}{c}{0.2, 0.252, 0.3175, 0.4, 0.504, 0.635, 0.8, 1.0079, 1.2699, 1.6 $\mu$m} \\
            \midrule
        Mie/DHS & \multicolumn{3}{c}{0.1 ,0.12 ,0.16, 0.20, 0.26, 0.34,  0.43, 0.55, 0.69, 0.89, 1.13, 1.44, 1.83,  2.34, 2.98, 3.79, 4.83, 6.16, 7.85, 10. $\mu$m}\\
        \midrule
    \end{tabular}}
    \tablefoot{For aggregates, we list the number of monomers linked to the maximum size of the aggregate. The size distribution is calculated by fixing the minimum grain size to the smallest available. These dust shape models must be considered in both possible compositions (amc or org), as with the 20 spherical models using Mie theory or DHS.}
    \label{tab:dustmodels}
\end{table*}

\vspace{-0.5cm}
\section{Impact of the disk geometry on color}
\subsection{Inclination and disk scale height} \label{APP:Geometry}
\begin{figure}[h!]
    \centering
    \includegraphics[width=0.9\linewidth]{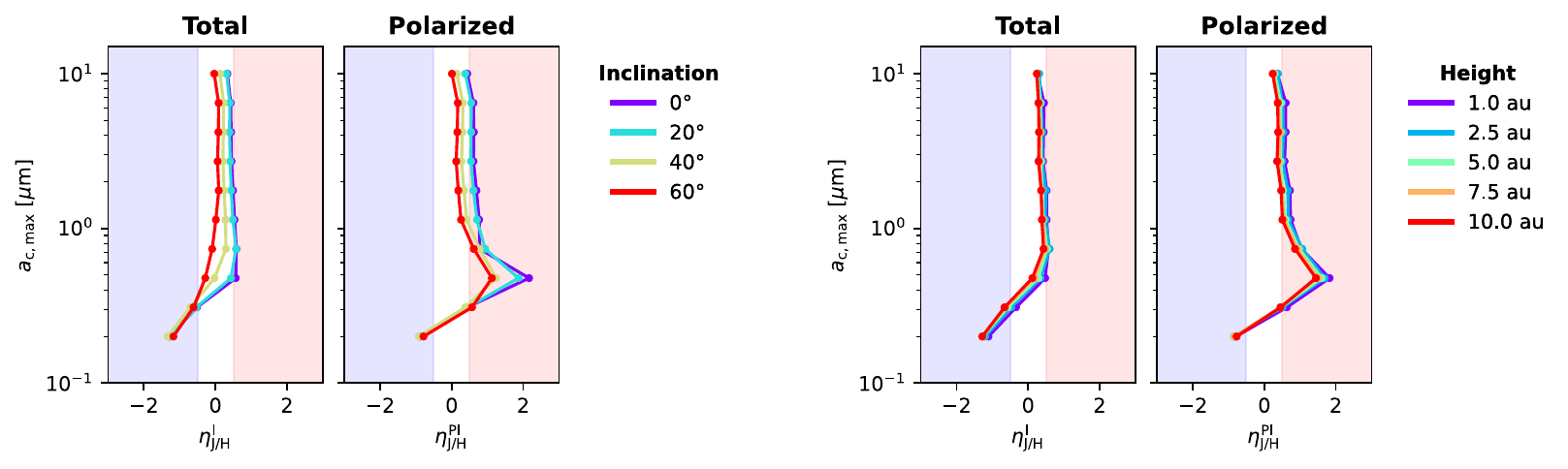}
    \caption{Impact of disk inclination (\textit{left}) and gas pressure scale height (\textit{right}) on the color $\eta_\mathrm{J/H}$ in total and polarized intensity.}
    \label{fig:geometry}
\end{figure}

In this section we verify that the measurement of the colors of protoplanetary disks is weakly dependent on their geometry. We used a disk composed of amc spherical grains following the Mie theory, with the same size distribution as before. We varied the disk inclination from $0^\circ$~to $60^\circ$, and the gas pressure scale height from $h_0=1$ to $10$~au at a reference radius of $r_0=100$~au. Modifying this gas pressure scale height also affects the apparent scattering surface height of the disk. The disk color, displayed in figure \ref{fig:geometry}, shows that the scattering surface height does not strongly affect the disk color.

For different inclinations, the disk color is slightly bluer when the inclination increases, for both $\eta^\mathrm{I}$ and $\eta^\mathrm{PI}$. In polarized intensity, the bump toward the red seen at $\sim 0.5\mu$m decreases when the inclination increases. These observations can be slightly affected by the composition of the dust within the disk. We measure shifts of $\Delta \eta \approx 0.1$ and $\Delta \eta \approx 0.01$ between disk models that vary by $20^\circ$ in inclination and $2.5~$au in scale height, respectively. Therefore, we can assume that the impact of disk geometry on the color observed between the J and H bands ($1.25~\mu$m and $1.63~\mu$m) remains negligible and does not strongly alter the color trends as a function of the dust grain size.

\vspace{-0.5cm}
\subsection{Impact of the inner disk} \label{APP:InnerDisk}
\begin{figure}[h!]
    \centering
    \includegraphics[width=0.9\linewidth]{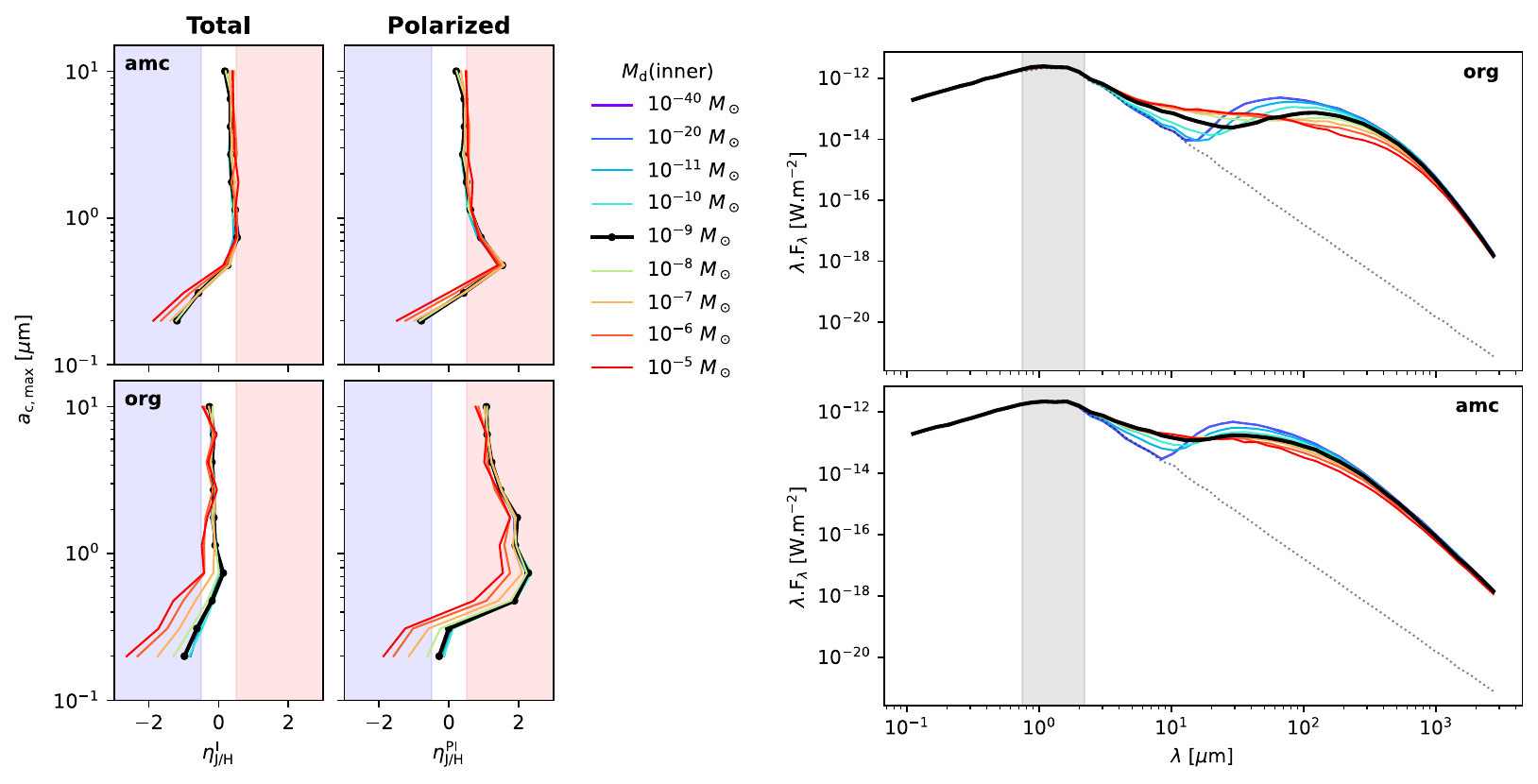}
    \caption{Effect of the mass of the inner disk on the color $\eta_\mathrm{J/H}$ in total and polarized intensity (\textit{left}), and on the SED (\textit{right}), for spherical grains with compositions amc (\textit{top}) and org (\textit{bottom}). The photosphere component in the SED is shown as a dotted gray line.}
    \label{fig:InnerDisk}
\end{figure}

In this section we investigate the impact of the presence of an inner disk in our model. We used the same disk model as described in Sect. \ref{Sect:Diskmodel}, varying only the dust mass of the inner disk, up to $M_\mathrm{d, inner}=10^{-5}M_\odot$ in the limiting case of an extremely massive inner disk. The inner disk will progressively become denser, and this may affect color measurements. The disk model is composed of spherical grains following Mie theory, and we considered both amc and org compositions. Figure \ref{fig:InnerDisk} shows the measurement of color $\eta_\mathrm{J/H}$ with respect to $a_\mathrm{c, max}$ in the left panel, and the corresponding spectral energy distribution (SED) in the right panel.

In the case of an amc composition, the presence of an inner disk has almost no impact on the disk color in both total and polarized intensity. However, for org composition, the inner disk starts to affect the color for a mass above $10^{-7}M_\odot$, where \etaI~and \etaPI~ tend to shift toward the blue, particularly for the small grains. The outer disk receives light that has already been forward-scattered by the dust grains within the inner disk surface. Moreover, the effect of the inner disk is largely visible for more reflective dust composition (org), for which multiple scattering is favored and the contribution of forward scattering by dust grains from the inner disk surface increases. SEDs show that an increase in inner disk mass increases the excess in the near-infrared, but the flux between 0.554 and 2.18~$\mu$m remains dominated by the stellar photosphere. The thermal emission from the inner disk therefore does not significantly contribute to the wavelengths used to measure colors. The color changes seen for org composition at high inner disk masses ($\geq10^{-7} M_\odot$) are thus due to multiple scattering and shadowing by the inner disk, not to thermal emission.

The dust mass of the outer disk located between $50$ and $200$~au, is set to $10^{-5}M_\odot$. The surface density in both inner and outer disks is defined as $\Sigma(r) \propto r^{-0.5}$. If we extend the dust mass from the outer to the inner disk  with this law, it implies a typical dust mass of approximately $10^{-8}M_\odot$ for the inner disk. Given that previous studies have estimated a mass of the inner disk ranging from $10^{-11} M_\odot$ to $10^{-9} M_\odot$ in some protoplanetary disks \citep{Olofsson2013, Perez2015, Villenave2019}. Our standard model assuming an inner disk with $M_d = 10^{-9} M_\odot$ is representative of such observations. Given these, an inner disk with a dust mass larger than $10^{-7} M_\odot$ seems very unlikely. Looking only at coherent masses, we can conclude that the effect of an inner disk is negligible for the measurements of disk color.

\vspace{-0.5cm}
\section{All colors} \label{APP:AllColor}
\begin{figure}[p]
    \centering
    \includegraphics[width=0.95\linewidth]{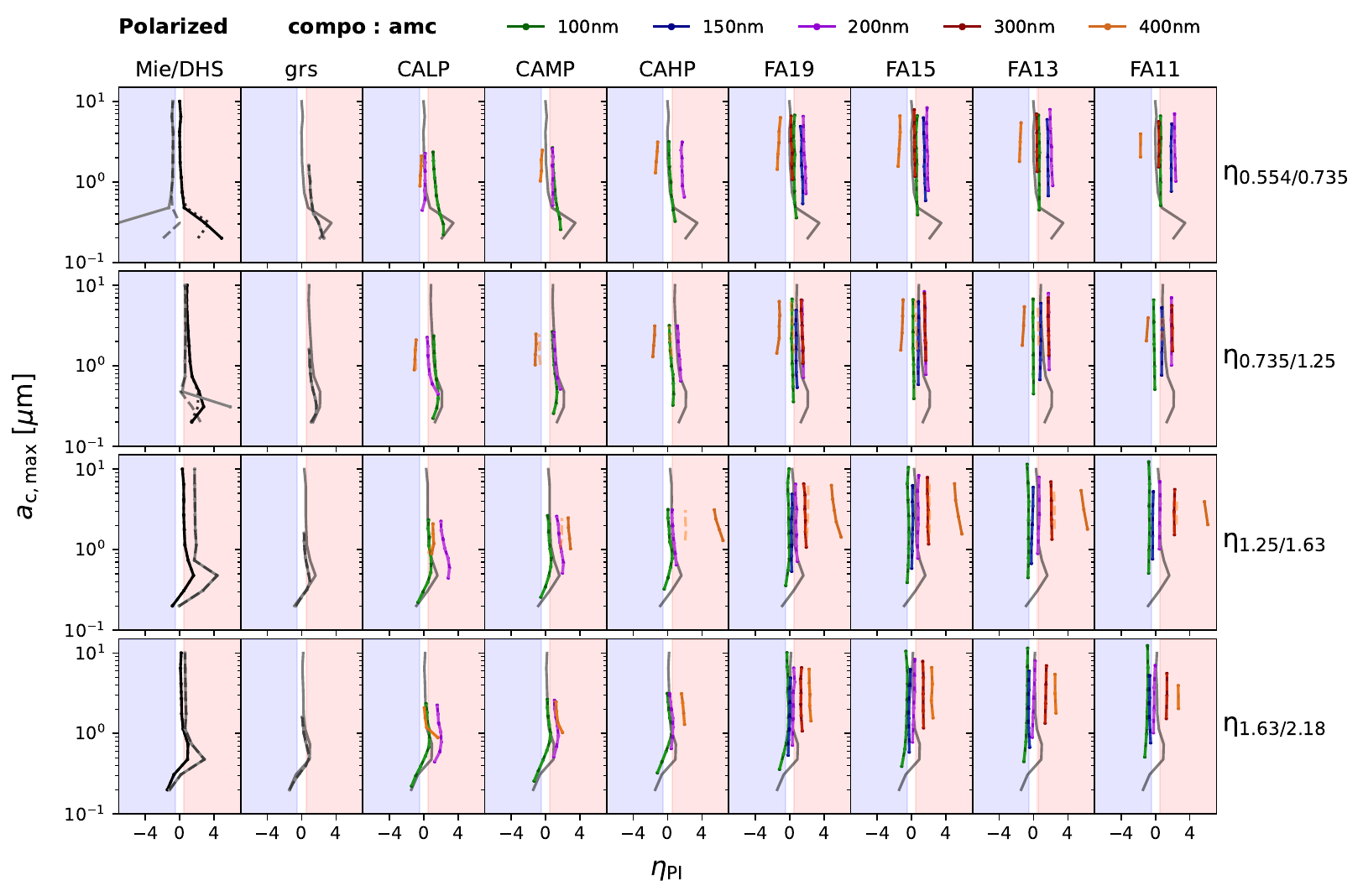}\\
    \includegraphics[width=0.95\linewidth]{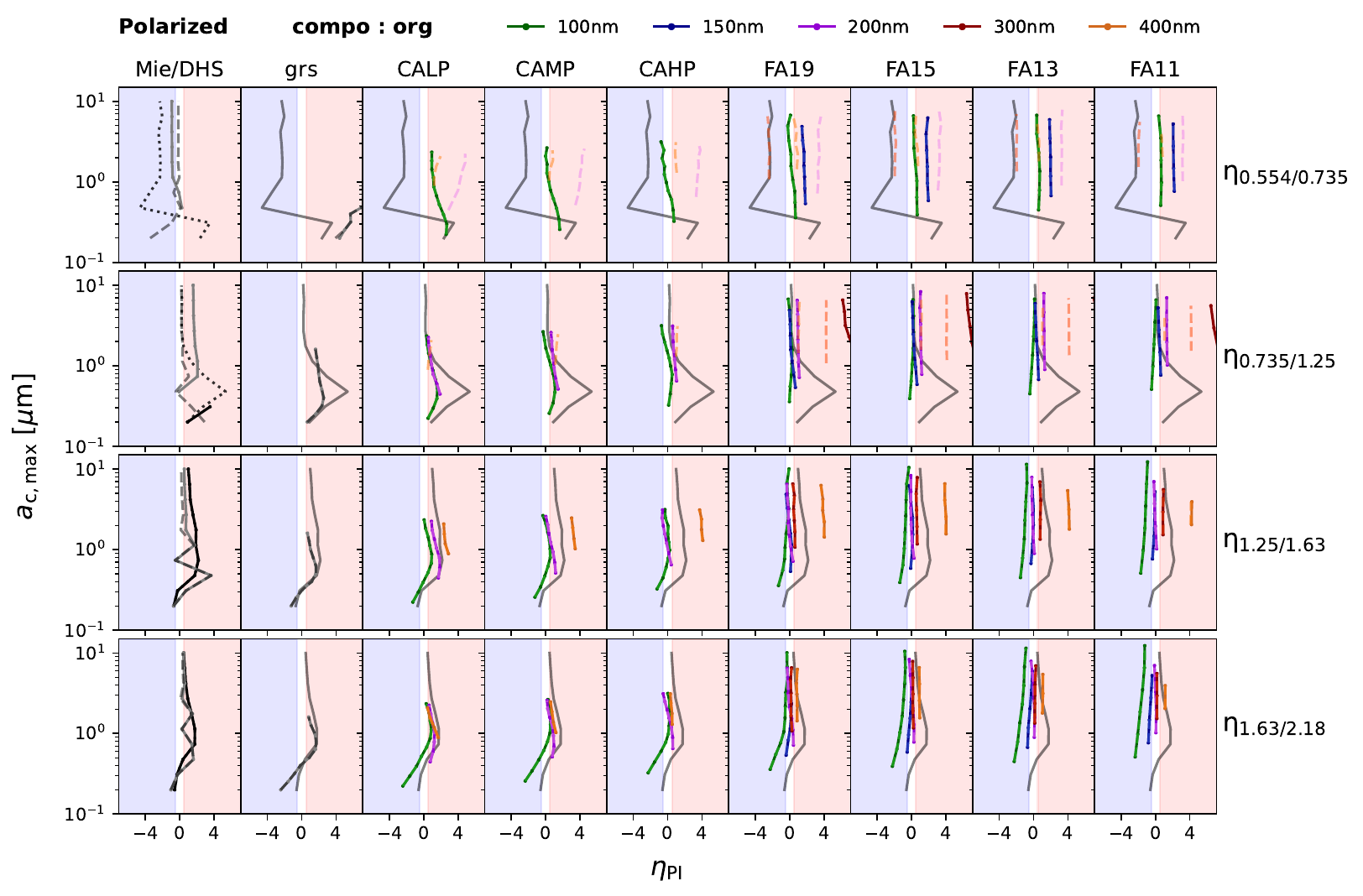}\\
    \caption{Polarized disk colors between consecutive wavelengths from $0.554$ to $2.18~\mu$m, as a function of grain size, for amc (\textit{top}) and org (\textit{bottom}) compositions. Colors computed from $Q_\phi$ images are shown as solid curves with dots, and those from $\lvert Q_\phi\rvert$ as dotted curves. Each column corresponds to one particle type, from compact spherical grains (Mie/DHS; \textit{left}) to fractal aggregates (\textit{right}). For complex particles, the Mie colors are overplotted in gray for comparison. For aggregates, the colors are separated by monomer size.}
    \label{Appfig:PolarizedColor}
\end{figure}

\begin{figure}[p]
    \centering
    \includegraphics[width=0.95\linewidth]{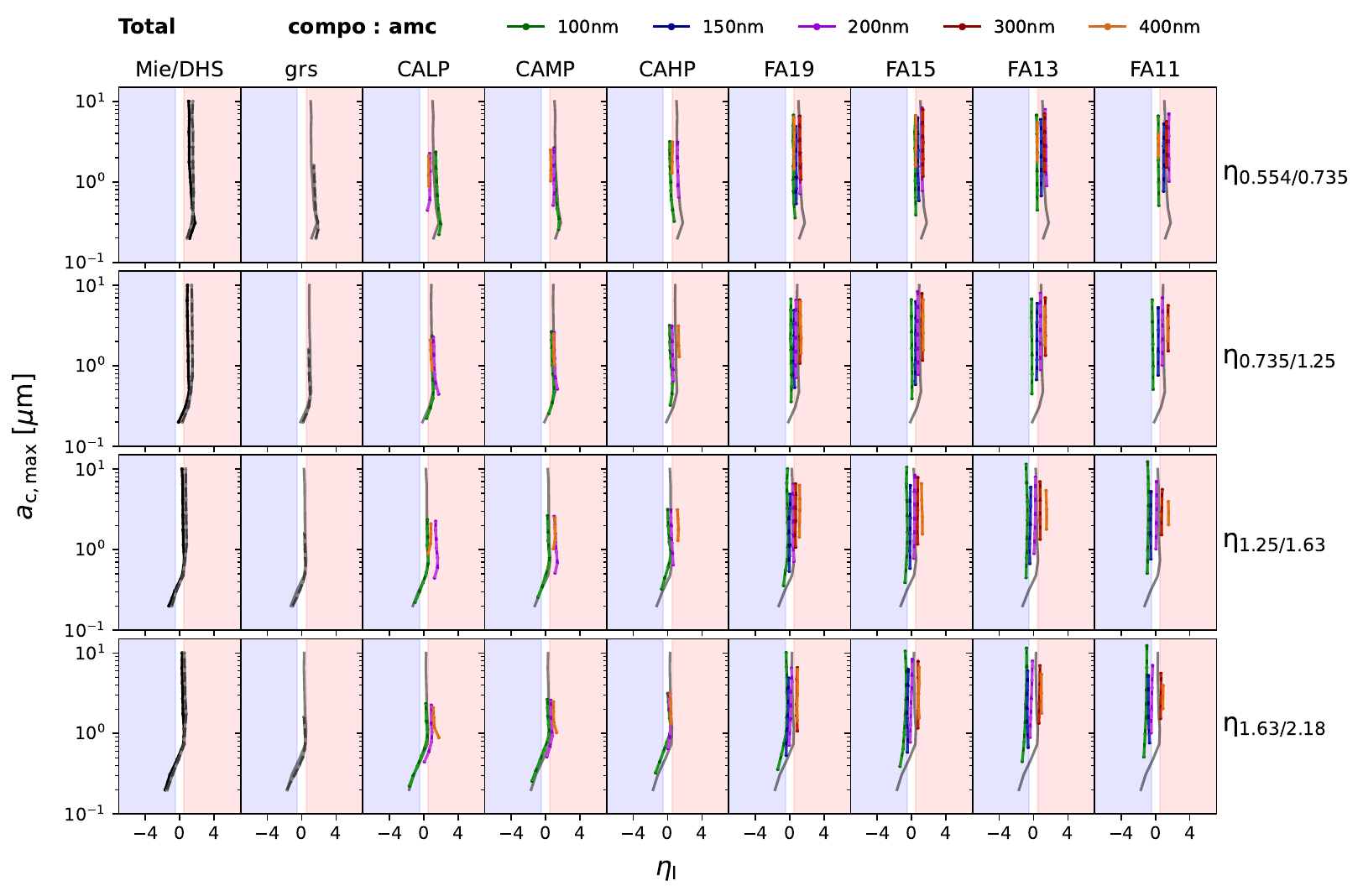}\\
    \includegraphics[width=0.95\linewidth]{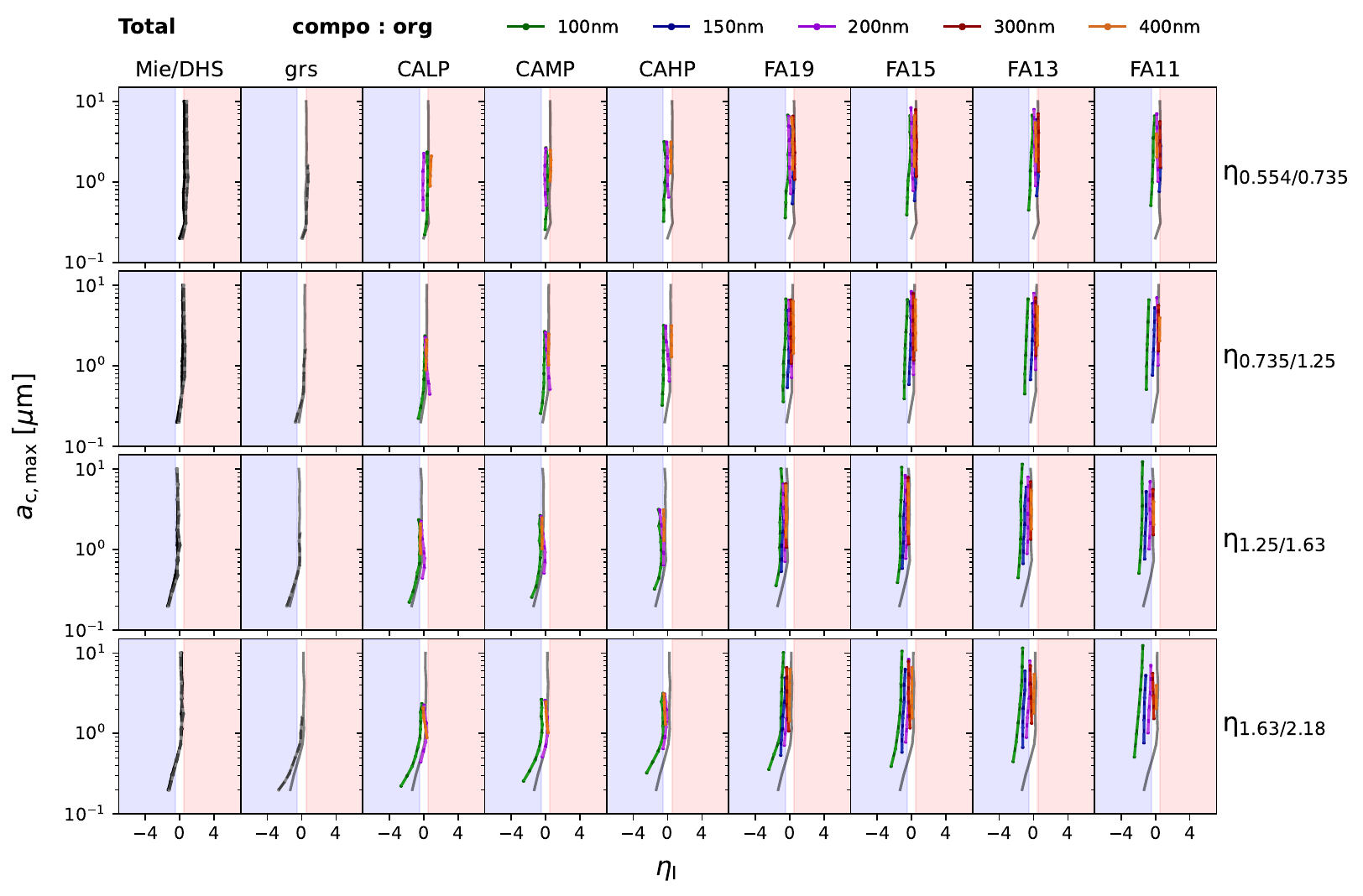}\\
    \caption{Same as Fig. \ref{Appfig:PolarizedColor} but for total intensity}
    \label{Appfig:TotalColor}
\end{figure}

In this section we present all the total and polarized colors obtained for $\lambda = 0.554~\mu$m, $0.735~\mu$m, $1.25~\mu$m, $1.63~\mu$m, and $2.18~\mu$m, for all the dust particle types, from spherical grains following the Mie theory to fractal aggregates \citep{tazaki_size_2022, tazaki_fractal_2023}. All measurements are made on synthetic observations of protoplanetary disks, considering a particle size distribution (see Figs. \ref{Appfig:PolarizedColor} and \ref{Appfig:TotalColor}). This supplementary analysis highlights the importance of multiwavelength studies, and especially those involving polarized intensity, to better understand the properties and evolution of dust in protoplanetary disks.

Figure~\ref{Appfig:PolarizedColor} also shows a comparison between the polarized colors measured using $Q_\phi$ and or $\lvert Q_\phi\rvert$. We find that the two measurements can differ, especially for $\eta_\mathrm{0.554/0.735}$ and $\eta_\mathrm{0.735/1.25}$ when considering dust model with org composition, compact spherical grains, and aggregates with large monomers. These differences occur when the synthetic images of $Q_\phi$ contain negative flux, which is uncommon in telescope observations. The negative values of $Q_\phi$ correspond to a negative DoLP, $-S_{12}/S_{11} < 0$, which occurs at certain scattering angles depending on the composition, size, and size distribution of the particles. For compact spherical grains (Mie theory) with org composition, $Q_\phi$ images at 0.554 and 0.735~$\mu$m are entirely negative: even the smallest particles in the distribution ($0.2~\mu$m) have a size parameter $x = 2\pi a/\lambda \gtrsim 1$ and are therefore beyond the Rayleigh regime. For aggregates, polarization is governed by the size of the monomers, and aggregate models assume monodisperse monomers (single size). This can produce strong oscillations, or even a sign change in the DoLP as a function of the scattering angle. 
When computing the color from $Q_\phi$, the positive and negative pixels cancel each other out, whereas $\lvert Q_\phi\rvert$ adds them together, resulting in the discrepancy between the two colors. These negative fluxes therefore reflect the limitations of our dust models rather than a property of the observed disks and we expect that most regions of strong negative polarization will disappear when a monomer size distribution is taken into account. Then, when a negative polarization branch appears in the relevant scattering angles probed by the disk, the measured flux, and thus the color, depends on whether $Q_\phi/I_\star$ or $\lvert Q_\phi\rvert/I_\star$ is used to compute the polarized intensity color.

\end{document}